\documentclass[prd,twocolumn,eqsecnum,showpacs,amsfonts,amssymb]{revtex4}

\usepackage{graphicx}

\usepackage{bm}

\newcommand{\beq}{\begin{equation}}
\newcommand{\eeq}{\end{equation}}
\newcommand{\beqs}{\begin{eqnarray}}
\newcommand{\eeqs}{\end{eqnarray}}
\newcommand{\lsim}{\mathrel{\raisebox{-
.6ex}{$\stackrel{\textstyle<}{\sim}$}}}

\begin{document}

\title{On the Question of an Ultraviolet Zero of the Beta Function of the 
$\lambda (\vec{\phi}^{\, 2})^2_4$  Theory}

\author{Robert Shrock}

\affiliation{C. N. Yang Institute for Theoretical Physics \\
Stony Brook University \\
Stony Brook, NY 11794 }

\begin{abstract}

We investigate the possibility of an ultraviolet (UV) zero in the $n$-loop beta
function of a $\lambda (\vec{\phi}^{\, 2})^2$ field theory with an
$N$-component scalar field, $\vec{\phi}$, in four spacetime
dimensions, up to the level of $n=5$ loops. Although the two-loop beta function
has a UV zero, we find that the $n$-loop beta function
for $n=3, \ 4, \ 5$ either does not have a UV zero or does not have one at a
value of $\lambda$ in approximate agreement with the two-loop calculation.
Similar results are obtained after application of scheme transformations to
the beta function and via calculation of Pad\'e approximants. We thus conclude
that in the range of $\lambda$ where the perturbative calculation of the 
$n$-loop beta function is reliable, the theory does not exhibit robust
evidence of a UV zero up to the level of $n=5$ loops. 

\end{abstract}

\pacs{11.10.-z,,11.10.Hi}

\maketitle

% =======================================================================

\section{Introduction}
\label{intro}

A subject of fundamental interest in quantum field theory is the dependence of
the interaction coupling on the Euclidean momentum scale, $\mu$,
where it is measured.  This dependence is described by the beta function of the
theory \cite{rg}.  Here we investigate this for a theory with a real,
$N$-component, scalar field, $\vec{\phi} = (\phi_1,...,\phi_N)^T$ with a
self-interaction of the form $\lambda (\vec{\phi}^{\, 2})^2$, in $d=4$
spacetime dimensions (at zero temperature), focusing on the question of whether
the beta function exhibits robust evidence of an ultraviolet zero. We
study this up to the highest loop order for which the beta function has been
calculated, namely five loops.  This theory is defined by the path integral
\beq
Z = \int \prod_x [d \vec{\phi}(x)] \, e^{iS} \ , 
\label{zphi}
\eeq
where $S=\int d^4x \, {\cal L}$, with the Lagrangian \cite{alt}
\beq
{\cal L} = \frac{1}{2}(\partial_\mu \vec{\phi}) \cdot
                      (\partial^\mu \vec{\phi})
- \frac{m^2}{2}\vec{\phi}^{ \, 2} - 
\frac{\lambda}{4!}(\vec{\phi}^{\, 2})^2 \ . 
\label{lag}
\eeq
This theory will be denoted $\lambda (\vec{\phi}^{\, 2})^2_4 = 
\lambda |\vec{\phi}|^4_4$ for short, where the subscript means $d=4$. 
The Lagrangian ${\cal L}$ is invariant under global 
O($N$) transformations of the field $\vec{\phi}$.  Quantum loop corrections
lead to a scale dependence in the physical, renormalized coupling,
$\lambda(\mu)$.  

The variation of $\lambda(\mu)$ as a function of $\mu$ is described by the
beta function
\beq
\beta_\lambda = \frac{d\lambda}{dt} \ ,
\label{betadef}
\eeq
where $dt=d \ln \mu$. (The argument $\mu$ will often be suppressed in the
notation.)  As is well-known, the lowest-order (one-loop) term in this beta
function has a positive coefficient, so that as $\mu \to 0$, the coupling
$\lambda(\mu) \to 0$, i.e., the theory is infrared-free.  This perturbative
result has been confirmed by nonperturbative approaches \cite{nonpert} and is
sometimes referred to as the ``triviality'' property of the theory.  One then
interprets the theory as an effective one, applicable over some range of
momenta $\mu$ (see, e.g., \cite{weinbergbook,zinnjustinbook}).  Since the
one-loop term in $\beta_\lambda$ is positive, it follows that as $\mu$
increases from 0, the coupling $\lambda(\mu)$ also increases.  If one were
naively to consider only the lowest-order term in the beta function and
integrate the differential equation (\ref{betadef}), then this would lead to a
pole in $\lambda(\mu)$ at a finite value of $\mu$. Of course, one would not
actually be justified in drawing such an inference, since as $\mu$ increased,
$\lambda(\mu)$ would become too large for the perturbative calculation to be
valid before the position of the pole would be reached.  However, this
motivates one to consider higher-loop terms in the beta function.

An important question is whether $\beta_\lambda$ has a UV zero, which could
thus constitute an ultraviolet fixed point (UVFP) of the renormalization group
(RG), so that as $\mu$ increases from the infrared (IR) limit $\mu=0$ to the UV
limit $\mu \to \infty$, $\lambda(\mu)$ would increase, but approach a finite
value.  In this paper we shall investigate this question using higher-loop
calculations of the beta function up to five-loop order.  The two-loop term in
$\beta_\lambda$ is negative, so that, at the two-loop level, $\beta_\lambda$
does, in fact, exhibit a UV zero.  If the existence of this zero were to be
confirmed at higher loop order, then it would mean that the growth of
$\lambda(\mu)$ would be cut off for large $\mu$.  Although the $\lambda
|\vec{\phi}|^4_4$ theory has been studied for many years, we are not aware of a
paper in the literature that has addressed and answered the question of whether
the higher-loop beta function also exhibits robust evidence for a UV zero. A
prerequisite for a UV zero of $\beta_\lambda$ to be well-established, at least
within the context of a perturbative calculation, is that for a given value of
$N$, when one calculates the value of the zero to $n$-loop order and to
$(n+1)$-loop order, there should not be a large fractional shift in the
result. We shall investigate this question both for a range of finite values of
$N$ and in the large-$N$ limit.  Early discussions of the large-$N$ limit of
the O($N$) $\lambda |\vec{\phi}|^4_4$ theory include \cite{aks}; see also
\cite{zinnjustinbook}. To state our conclusion at the outset, we find evidence
against a UV zero in this theory. This finding is consistent with the view of
this theory as an effective field theory, to be applied only over a restricted
range of momentum scales $\mu$.  What it contributes to this consensus is a
quantitative analysis of the perturbatively calculated $n$-loop beta function
up to the rather high order of five loops.

For our study of the ultraviolet properties of the beta function for large
momentum scale $\mu$, we take the quartic coupling $\lambda$ in ${\cal L}$ to
be positive for the stability of the theory.  Corrections to the scalar
potential have been discussed, e.g., in \cite{cw}.  We take the coefficient,
$m^2$, of the term $(1/2)\vec{\phi}^{\ 2}$ in ${\cal L}$ to be fixed and recall
that its value does not enter in the beta function $\beta_\lambda$ that we
analyze. If one were to consider this scalar theory as being embedded in a
larger theory with higher physical mass scales, then one would have to deal
with the hierarchy problem, namely the sensitivity, via loop corrections, of
$m^2$ to these higher mass scales.  To focus on the question of a UV zero of
the beta function, which is of interest in its own right, we thus consider this
theory in isolation.

The question of the existence of a UV zero in the beta function of the O($N$) 
$\lambda |\vec{\phi}|^4_4$ is somewhat similar to the question of the existence
of a UV zero in the beta function of a U(1) gauge theory in $d=4$ dimensions 
with a set of $N_f$ fermions of a given charge.  Both of these theories are
IR-free. In \cite{lnf} we recently investigated the question for the U(1) gauge
theory up to the five loop level for general $N_f$ and in the limit of large
$N_f$ and concluded that it does not exhibit a UV zero \cite{lnf} (see also
\cite{holdom}).  

An example of a theory with a beta function that has a UV zero is the
nonlinear $\sigma$ model (NL$\sigma$M) in $d=2+\epsilon$ spacetime dimensions,
where one finds, from an exact solution of this model in the limit $N \to
\infty$ (involving a sum of an infinite number of Feynman diagrams that
dominate in this limit) the result (for small $\epsilon$) \cite{nlsm}
\beq
\beta_\lambda = \epsilon \lambda \Big ( 1 - \frac{\lambda}{\lambda_c} \Big )
\ ,
\label{beta_nlsm}
\eeq
where $\lambda$ is an effective dimensionless coupling in the model and
\beq
\lambda_c = \frac{2\pi \epsilon}{N} \ . 
\label{lambda_c_nlsm}
\eeq
In addition to being an IR-free theory with a UVFP, this was also an early
example of a theory which, by perturbative power-counting for $\epsilon > 0$,
is nonrenormalizable, but nevertheless yields calculable, well-defined
predictions via the use of a nonperturbative method, namely the large-$N$
limit. This possibility has been termed ``asymptotic safety''
\cite{asymptotic_safety,weinbergbook}.

This paper is organized as follows.  In Sect. II we analyze the behavior of the
$n$-loop coefficients of the beta function of the $\lambda |\vec{\phi}|^4_4$
theory.  In Sect. III we investigate the question of the presence or absence of
a UV zero of the beta function up to five-loop order.  Sect. IV contains a
discussion of the large-$N$ limit.  In Sect. V we study the effect of applying
scheme transformations to the beta function in the analysis of a possible UV
zero.  Sect. VI contains a corresponding study using Pad\'e approximants.  In
Sect. VII we compare our findings with those for some related theories.  Our
conclusions are summarized in Sect. VIII. Some relevant formulas for
discriminants are given in the Appendix. Although we restrict our study here to
the beta function of the O($N$) $\lambda |\vec{\phi}|^4$ theory in $d=4$
dimensions, we note parenthetically that this field theory has
been used extensively to study critical phenomena in $d=4-\epsilon'$
dimensions, with particular application to $d=3$, and there has also been
interest in the $\lambda (\vec{\phi}^{ \ 2})^3$ theory in $d=3$ dimensions for
the study of tricritical points.

% ========================================================================

\section{Beta Function and Properties of Coefficients up to Five Loops}
\label{betafunction}

\subsection{General} 

The beta function $\beta_\lambda$ of Eq. (\ref{betadef}) 
has the series expansion
\beq
\beta_\lambda = \lambda \sum_{\ell=1}^\infty b_\ell \, a ^\ell \ , 
\label{beta}
\eeq
where
\beq
a \equiv \frac{\lambda}{16\pi^2} \ . 
\label{a} 
\eeq
An equivalent beta function is $\beta_a = da/dt$, with the series expansion
\beq
\beta_a = a \sum_{\ell=1}^\infty b_\ell \, a ^\ell \ . 
\label{betaa}
\eeq
The $n$-loop $\beta_\lambda$ and $\beta_a$ functions, 
denoted $\beta_{\lambda,n\ell}$ and $\beta_{a,n\ell}$, are 
given by Eqs. (\ref{beta}) and (\ref{betaa}) with
the upper limit of the loop summation index $\ell=n$ instead of $\ell=\infty$.
It will be convenient to define the scaled coefficients 
\beq
\bar b_\ell \equiv \frac{b_\ell}{(4\pi)^\ell} 
\label{bellbar}
\eeq
for tables to be presented later.  In our analysis of how the inclusion of
higher loops changes the value of the beta function for a given $a=a(\mu)$ and
$N$, it is useful to define the ratio
\beq
R_{a,n} \equiv \frac{\beta_{a,n\ell}}{\beta_{a,1\ell}} = 
1 + \sum_{\ell=2}^n \Big ( \frac{b_\ell}{b_1} \Big ) \, a^{\ell-1}  \ . 
\label{ran}
\eeq
%

% ======================================================================

\subsection{$b_1$ and $b_2$} 

The one-loop and two-loop coefficients in the beta function, $b_1$
and $b_2$, are independent of the scheme used for regularization and
renormalization, while the coefficients at loop order three and higher,
$b_\ell$ for $\ell \ge 3$, are scheme-dependent.  The first two coefficients 
are \cite{bgz74}
\beq
b_1 = \frac{1}{3}(N+8) 
\label{b1}
\eeq
and
\beq
b_2 = -\frac{1}{3}(3N+14) \ . 
\label{b2}
\eeq
As noted above, since $b_1 > 0$, it follows that for small $a$, where the
calculation of $\beta_a$ is most reliable, $\beta_a > 0$, so that as $\mu \to
0$, $a(\mu) \to 0$, i.e., the theory is IR-free.  Going in the opposite
direction, to large $\mu$ in the ultraviolet, $a(\mu)$ increases.  A basic
question is whether this growth is cut off by a UV zero of the beta function,
so that as $\mu \to \infty$, $a(\mu)$ approaches a fixed finite constant,
$a_{_{UV}} = \lambda_{UV}/(16\pi^2)$ or whether, in contrast, $\beta_a$ has no
(reliably calculable) UV zero, so that $a(\mu)$ continues to increase with
increasing $\mu$ until it exceeds the regime where the function $\beta_a$
describing its evolution can be reliably calculated.  As is evident from
Eq. (\ref{b2}, $b_2$ is negative, so the two-loop beta function does have an UV
zero.  However, one must study whether this is stable when higher-loop terms
are included in the beta function.  In order to carry out this analysis, we
first characterize the behavior of the higher-loop coefficients to the highest
order for which they have been calculated, namely $n=5$.

% ======================================================================

\subsection{$b_3$}

The convenient and widely used $\overline{MS}$ scheme employs dimensional
regularization \cite{dimreg} with modified minimal subtraction
\cite{msbar}.  In the $\overline{\rm MS}$ scheme, the three-loop coefficient is
\cite{bgz74,b345}
\beq
b_3=\frac{11}{72}N^2+\bigg ( \frac{461}{108} + \frac{20\zeta(3)}{9} \bigg ) N 
+ \frac{370}{27} + \frac{88\zeta(3)}{9} \ . 
\label{b3}
\eeq
(See also \cite{kleinertbook} for a review and \cite{vkt} for the $N=1$ special
case of $b_3$ in this scheme).  Numerically,
\beq
b_3 = 0.15278 N^2 + 6.93976 N + 24.4571 \ . 
\label{b3num}
\eeq
Here and below, numerical quantities are listed to the indicated
floating-point accuracies.  For all physical $N$, this coefficient is positive
and is a monotonically increasing function of $N$. For later reference, we list
values of this and the other coefficients of the beta function, expressed in
terms of the conveniently rescaled quantities $\bar b_\ell$ defined in
Eq. (\ref{bellbar}), in Table \ref{btilde_nloop_values}.

% ======================================================================

\subsection{$b_4$} 

The four-loop coefficient in $\beta$, calculated in the $\overline{\rm MS}$
scheme, is \cite{b345}
\begin{widetext}
\beqs
b_4 & = & \frac{5}{3888}N^3+\bigg ( -\frac{395}{243} - \frac{14\zeta(3)}{9}
+ \frac{10\zeta(4)}{27} - \frac{80\zeta(5)}{81} \bigg ) N^2
+ \bigg (-\frac{10057}{486} - \frac{1528\zeta(3)}{81} + \frac{124\zeta(4)}{27}
- \frac{2200\zeta(5)}{81} \bigg ) N \cr\cr
& - & \frac{24581}{486} -
\frac{4664\zeta(3)}{81} + \frac{352\zeta(4)}{27} - \frac{2480\zeta(5)}{27} \ .
\label{b4}
\eeqs
\end{widetext}
Numerically,
\beqs
b_4 & = & (1.2860 \times 10^{-3})N^3 - 4.11865N^2 - 66.5621N \cr\cr
    & - & 200.92637 \ .
\label{b4num}
\eeqs
In contrast to the lower-order coefficients $b_\ell$ with $\ell=1,2,3$, $b_4$
is neither a monotonic function of $N$ nor of fixed sign.  At $N=1$,
$b_4=-271.606$ (equivalently, $\bar b_4 = -0.010892$), and as $N$ increases,
$b_4$ decreases through negative values.  This coefficient reaches a minimum
value (i.e., $-b_4$ reaches a maximum value) at the large number \cite{nreal}
$N = 2143.16$ and then increases, passing through zero and becoming positive
as $N$ increases through the value $N=N_{b4z}$, where
\beq
N_{b4z} = 3218.755 \ .
\label{Nb4z}
\eeq
This is the relevant one among the three roots of the cubic equation $b_4=0$
(the other two roots occur at negative, and hence unphysical, values of $N$).
Values of $\bar b_4$ are given in Table \ref{btilde_nloop_values}.

% =======================================================================

\subsection{$b_5$}

In the $\overline{\rm MS}$ scheme the five-loop coefficient is \cite{b345} 
\begin{widetext}
\beqs
b_5 & = & \bigg ( \frac{13}{62208} - \frac{\zeta(3)}{432} \bigg ) N^4
+ \bigg ( \frac{6289}{31104} + \frac{26\zeta(3)}{81} - \frac{2\zeta(3)^2}{27}
- \frac{7\zeta(4)}{24} + \frac{305\zeta(5)}{243} - \frac{25\zeta(6)}{81} 
\bigg ) N^3 \cr\cr
& + & 
\bigg ( \frac{50531}{3888} + \frac{8455\zeta(3)}{486} - \frac{59\zeta(3)^2}{81}
- \frac{347\zeta(4)}{54} + \frac{7466\zeta(5)}{243} - \frac{1775\zeta(6)}{243}
+ \frac{686\zeta(7)}{27} \bigg )N^2 \cr\cr
& + & \bigg ( \frac{103849}{972} + \frac{69035\zeta(3)}{486} +
\frac{446\zeta(3)^2}{81} - \frac{2383\zeta(4)}{54} + \frac{66986\zeta(5)}{243}
- \frac{7825\zeta(6)}{81} + 343\zeta(7) \bigg )N \cr\cr
& + & \frac{17158}{81} + \frac{27382\zeta(3)}{81} + \frac{1088\zeta(3)^2}{27}
- \frac{880\zeta(4)}{9} + \frac{55028\zeta(5)}{81} - \frac{6200\zeta(6)}{27}
+ \frac{25774\zeta(7)}{27} \ . 
\label{b5} 
\eeqs
Numerically, 
\beq
b_5 = -(2.57356 \times 10^{-3})N^4 + 1.152827 N^3 + 72.23315N^2 + 771.20866N 
+ 2003.97619 \ . 
\label{b5num}
\eeq
\end{widetext}
As $N$ increases from 1, this coefficient is initially positive and is an
increasing function of $N$, but it reaches a maximum at $N=374.02$ and then
decreases as $N$ passes this value.  As $N$ increases through the value
$N=N_{b5z}$, where
\beq
N_{b5z} = 504.74 \ , 
\label{Nb5z}
\eeq
$b_5$ decreases through zero to negative values, and remains negative for
larger $N$ (see Table \ref{btilde_nloop_values}). Here $N_{b5z}$ is the
relevant root of the quartic equation $b_5=0$ (the other three roots occur
occur at negative, and hence unphysical, values of $N$).  These sign changes in
$b_4$ and $b_5$ (as calculated in the $\overline{\rm MS}$ scheme) at large $N$
affect the behavior of the higher-loop $\beta$ function, as will be evident
from our analysis below.

 A comment is in order here concerning the generic size of higher-loop
coefficients.  The expansion (\ref{betaa}) is a specific example of a series
expansion of a generic quantity ${\cal O}$ in this theory, which can
be written as
\beq
{\cal O} = \sum_n c_{{\cal O},n} a^n \ , 
\label{oh}
\eeq
where $n$ denotes the loop order.  From asymptotic estimates, it has been
concluded that for $n \gg 1$, the coefficient $|c_{{\cal O},n}|$ grows
asymptotically dominantly as a factorial, $\sim n!$ (with additional factors
including $a^n n^b$, where $a$ and $b$ are constants)
\cite{asymptotic,zinnjustinbook}.  Since higher-order terms are
scheme-dependent, it is understood that this is the generic behavior. This
forms a basis for the proof that perturbative power series expansions in this
theory are not Taylor series expansions with finite radii of convergence, but
instead are only asymptotic expansions. In the present theory, it is noteworthy
that, as is evident from our analysis above and from Table
\ref{btilde_nloop_values}, there are values of $N$ for which the coefficients
$b_4$ and $b_5$, as calculated in the $\overline{\rm MS}$ scheme, have zeros.
For a value of $N$ where $b_n$ vanishes (while $b_{n-1}$ does not vanish), it
will obviously not be the case that $|b_n/b_{n-1}|$ exhibits the generic
large-order growth as a function of $n$. Of course, since the loop orders $n=4$
and $n=5$ are not $\gg 1$, these zeros are not inconsistent with the dominant
$n!$ growth in magnitude for $n \gg 1$.

% ========================================================================

\subsection{$n$-Loop Beta Function and Associated Ratio $R_n$}

\begin{figure}
  \begin{center}
    \includegraphics[height=8cm,width=6cm]{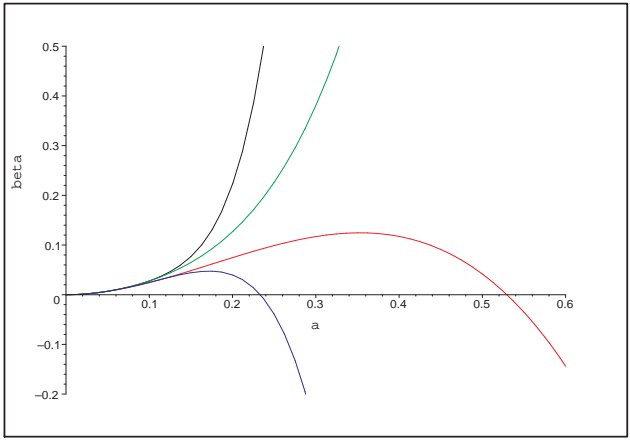}
  \end{center}
\caption{\footnotesize{ Plot of the $n$-loop $\beta$ function $\beta_{a,n\ell}$
as functions of $a$ for $N=1$ and (i) $n=2$ (red), (ii) $n=3$ (green), (iii)
$n=4$ (blue), and $n=5$ (black) (colors in online version). At $a=0.18$, going
from bottom to top, the curves are for $n=4$, $n=2$, $n=3$, and $n=5$.}}
\label{lam_beta_Neq1}
\end{figure}

\begin{figure}
  \begin{center}
    \includegraphics[height=8cm,width=6cm]{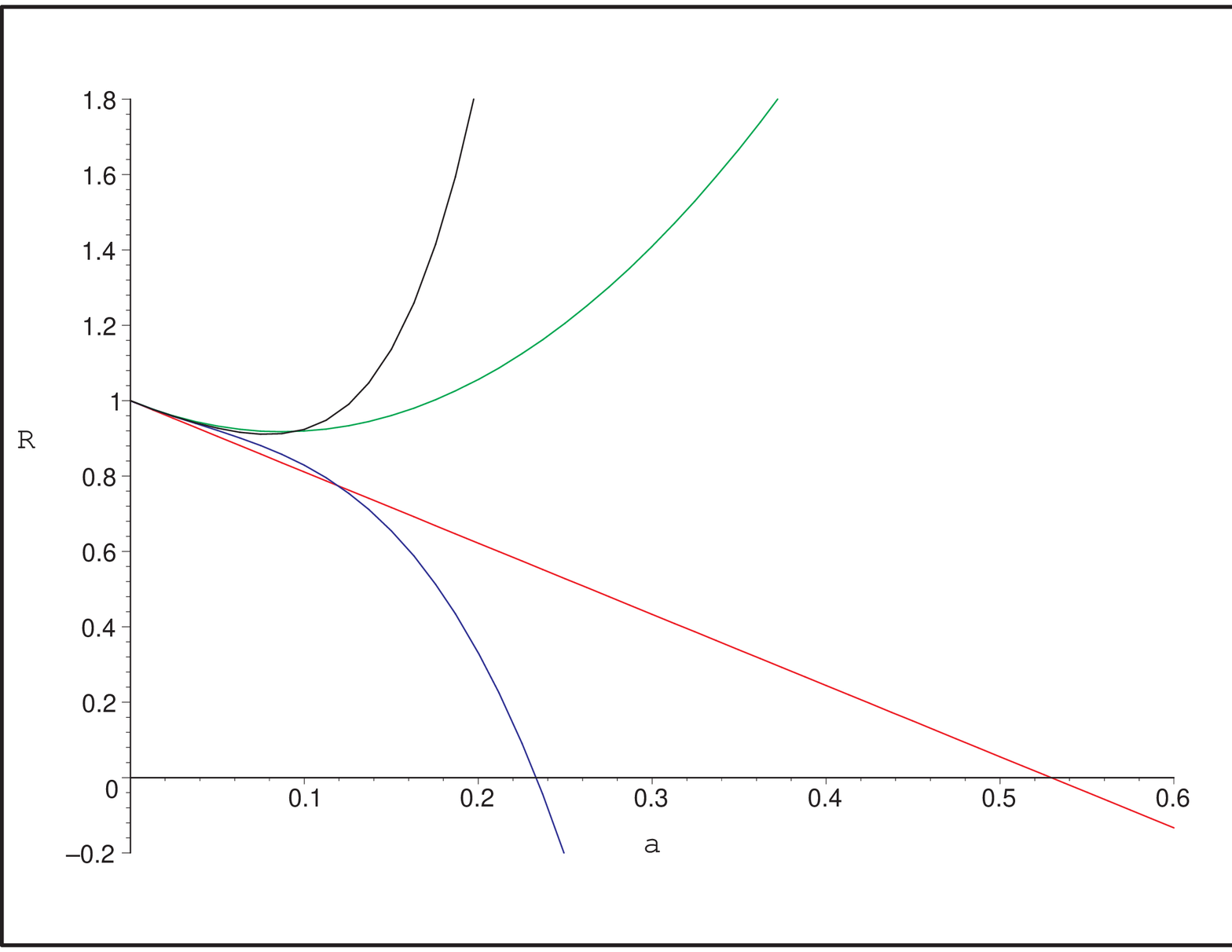}
  \end{center}                             
\caption{\footnotesize{Plot of the ratio $R \equiv R_{a,n}$ of
    $\beta_{a,n\ell}$ divided by $\beta_{a,1\ell}$, as a function of $a$ for
    $N=1$ and (i) $n=2$ (red), (ii) $n=3$ (green), (iii) $n=4$ (blue), and
    $n=5$ (black) (colors in online version). At $a=0.18$, going from bottom to
    top, the curves are for $n=4$, $n=2$, $n=3$, and $n=5$.}}
\label{lam_betareduced_Neq1}
\end{figure}

\begin{figure}
  \begin{center}
    \includegraphics[height=8cm,width=6cm]{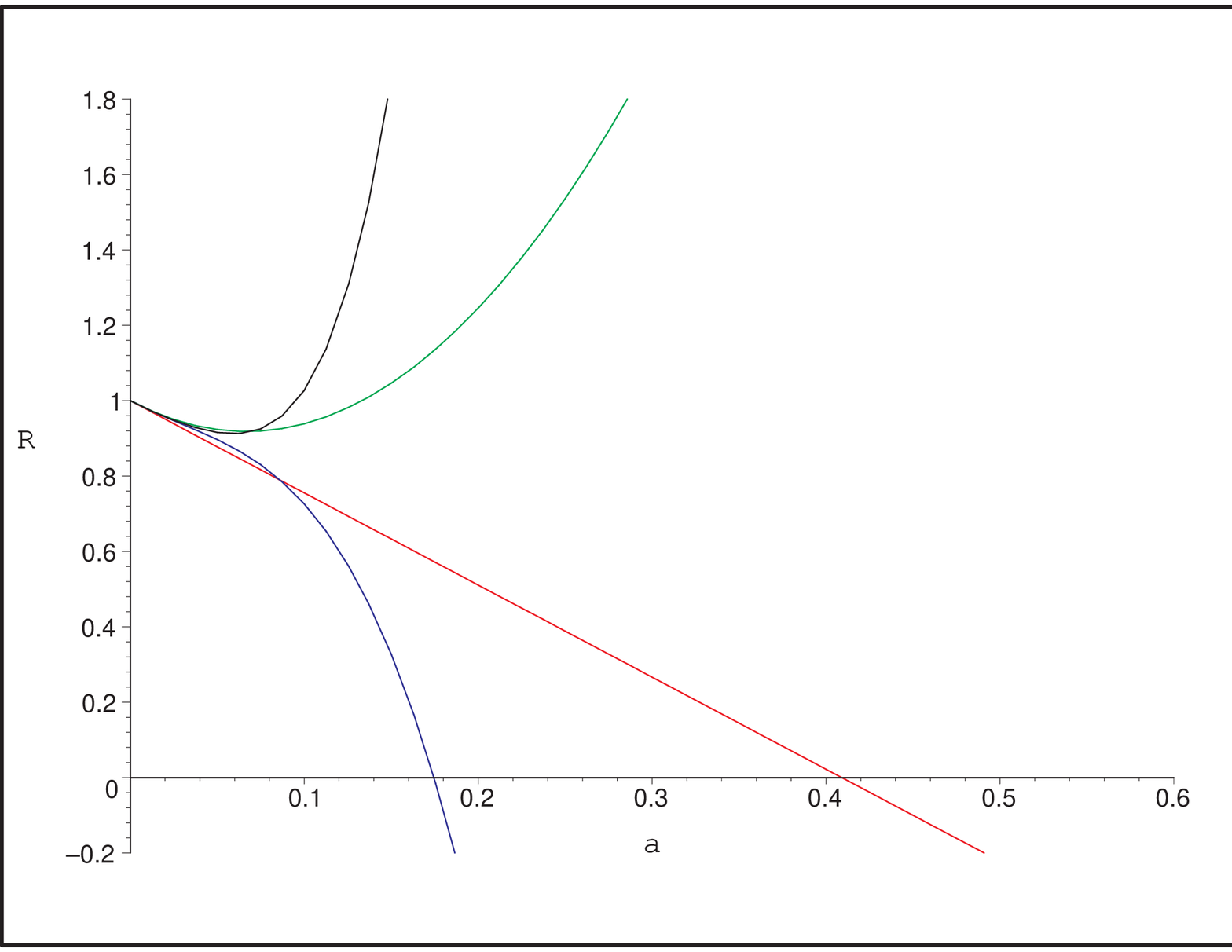}
  \end{center}                             
\caption{\footnotesize{Plot of the ratio $R \equiv R_{a,n}$ of
    $\beta_{a,n\ell}$ divided by $\beta_{a,1\ell}$, as a function of $a$ for
    $N=10$ and (i) $n=2$ (red), (ii) $n=3$ (green), (iii) $n=4$ (blue), and
    $n=5$ (black) (colors in online version).  At $a=0.14$, going from bottom
    to top, the curves are for $n=4$, $n=2$, $n=3$, and $n=5$.  }}
\label{lam_betareduced_Neq10}
\end{figure}

\begin{figure}
  \begin{center}
    \includegraphics[height=8cm,width=6cm]{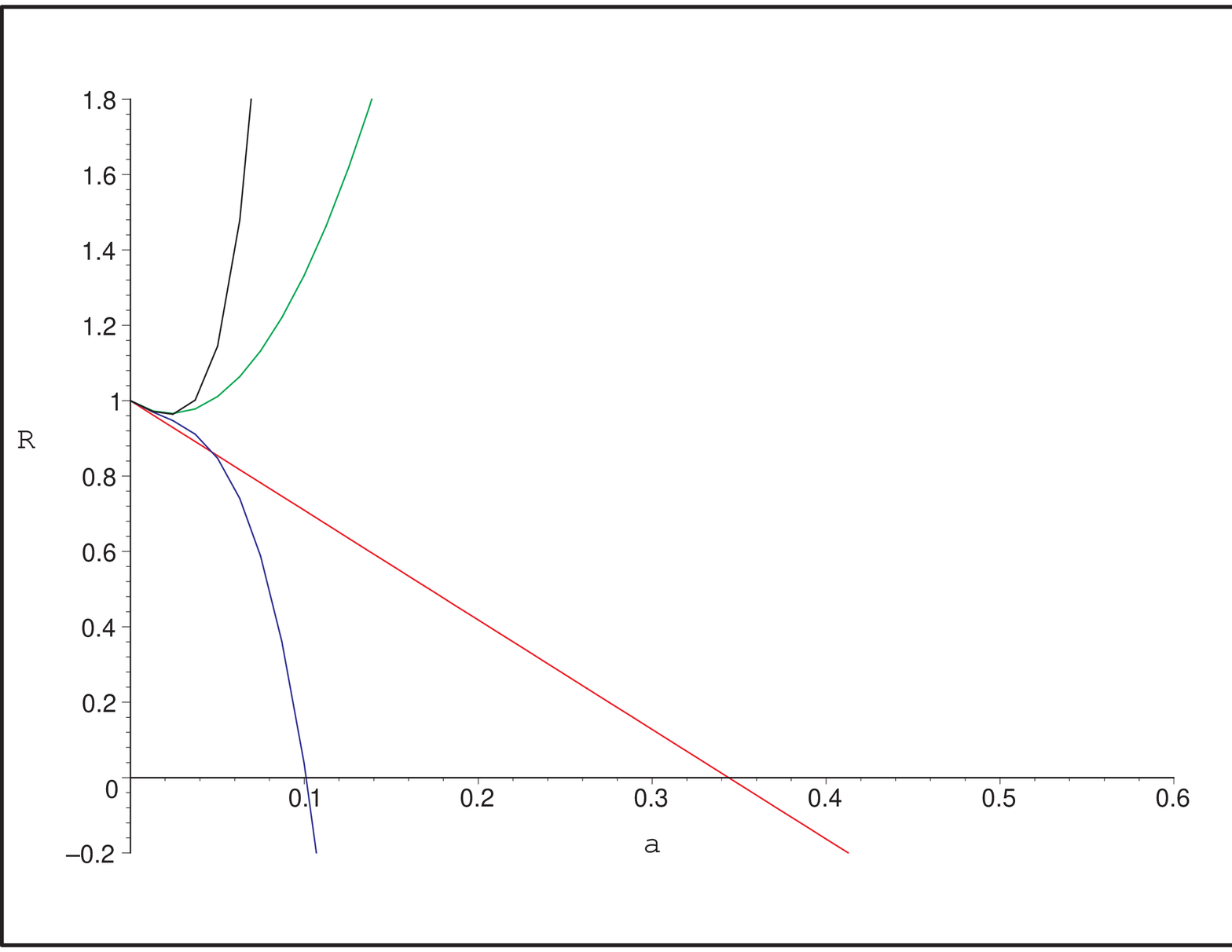}
  \end{center}
\caption{\footnotesize{Plot of the ratio $R \equiv R_{a,n}$ of
    $\beta_{a,n\ell}$ divided by $\beta_{a,1\ell}$, 
as a function of $a$ for $N=100$ and (i) $n=2$ (red), 
(ii) $n=3$ (green), (iii) $n=4$ (blue), and $n=5$ (black)
(colors in online version).
At $a=0.06$, going from bottom to top, the curves are for
$n=4$, $n=2$, $n=3$, and $n=5$.}}
\label{lam_betareduced_Neq100}
\end{figure}

In Fig. \ref{lam_beta_Neq1} we plot the respective $n$-loop beta functions
$\beta_{a,n\ell}$ for $n=2, \ 3, \ 4, \ 5$ loops and $N=1$. This plot shows the
ranges in $a$ over which the calculations of the beta function to various loop
orders agree with each other.  Another useful way of showing this is to plot
the ratio $R_{a,n}$ of $\beta_{a,n\ell}$ divided by $\beta_{a,1\ell}$, as
defined in Eq. (\ref{ran}), and we do this in Figs.
\ref{lam_betareduced_Neq1}-\ref{lam_betareduced_Neq100} for the cases $N=1$, 
$N=10$, and $N=100$, respectively. Clearly, 
\beq
\frac{R_{n+1}}{R_n} = \frac{\beta_{a,(n+1)\ell}}
                           {\beta_{a,n\ell}} \ , 
\label{rr}
\eeq
so that the ratio of ratios in Eq. (\ref{rr}) measures the extent to which the
$n$-loop and $(n+1)$-loop beta functions agree in value for a given $a$ and
$N$.

In the case $N=1$, as is evident from Figs.  \ref{lam_beta_Neq1} and
\ref{lam_betareduced_Neq1}, $\beta_{a,2\ell}$ and $\beta_{a,3\ell}$
(equivalently, the curves for $R_2$ and $R_3$) are close to each other in the
interval $0 \le a \lsim 0.04$, but as $a$ increases beyond 0.04,
$\beta_{a,3\ell}$ deviates progressively upward relative to
$\beta_{a,2\ell}$. At higher-loop order, $\beta_{a,3\ell}$ and
$\beta_{a,4\ell}$ (equivalently, the curves for $R_3$ and $R_4$) are close to
each other in essentially the same interval $0 \le a \lsim 0.04$, and for
larger $a$, $\beta_{a,4\ell}$ deviates below $\beta_{a,3\ell}$ (and eventually
also below $\beta_{a,2\ell}$).  Even going as high as five-loop order does not
significantly increase the interval in $a$ in which the beta functions
calculated to the highest two successive loop orders, namely $\beta_{a,4\ell}$
and $\beta_{a,5\ell}$ in this case, agree with each other.  Numerically, the
values of $\beta_{a,4\ell}$ and $\beta_{a,5\ell}$ (equivalently, the curves for
$R_4$ and $R_5$) are close to each other only for $a$ up to about 0.05.  For
larger $a$, $\beta_{a,5\ell}$ deviates progressively upward relative to
$\beta_{a,4\ell}$.  An important conclusion from this analysis and from Figs.
\ref{lam_beta_Neq1}-\ref{lam_betareduced_Neq100} is that the zero in the
two-loop beta function (for each of the values of $N$) occurs at too large a
value of $a$ for the perturbative calculation to be reliable. We have also
established the same result for other values of $N$. 

This behavior contrasts with the situation concerning an IR zero of the beta
function of an asymptotically free non-Abelian gauge theory when calculated to
progressively higher loop orders, up to loop order $n=4$
\cite{gk98}-\cite{lnn}.  For example, consider an SU($N_c$) gauge theory with
$N_f$ fermions in the fundamental representation.  As one can see from Fig. 1
in Ref. \cite{bc} for an SU(2) gauge theory with $N_f=8$ fermions and from
Fig. 2 in \cite{bc} for an SU(3) gauge theory with $N_f=12$ fermions, the range
of squared gauge couplings $\alpha = g^2/(4\pi)$ over which the three-loop and
four-loop beta functions for $\alpha$ agree with each other is significantly
larger and extends to stronger coupling than the range of $\alpha$ for which
the two-loop and three-loop beta functions are in close agreement. 

% ======================================================================

\section{Zeros of the Beta Function}

\subsection{$\beta_{a,2\ell}$} 

In this section we discuss the zeros of the $n$-loop $\beta$ function,
$\beta_{a,n\ell}$.  Clearly, $\beta_{a,n\ell}$ has a double zero at the origin.
In addition to the zero at $a=0$, as is well-known, the two-loop beta function,
$\beta_{a,2\ell}$, has a UV zero at $a=a_{_{UV,2\ell}}$, where
\beq
a_{_{UV,2\ell}} = -\frac{b_1}{b_2} = \frac{N+8}{3N+14} \ . 
\label{auv_2loop}
\eeq
This UV zero of $\beta_{a,2\ell}$ is a monotonically decreasing function
of $N$ for physical $N$, which decreases from the value
\beq
a_{_{UV,2\ell}} = \frac{9}{17} \quad {\rm at} \ N=1
\label{auv_2loop_Neq1}
\eeq
and approaches the limit 
\beq
\lim_{N \to \infty} a_{_{UV,2\ell}} = \frac{1}{3} \ . 
\label{auv_2loop_Neqinfty}
\eeq
As is evident from Eq. (\ref{auv_2loop}), the corresponding value of 
$\lambda_{UV,2\ell}$ is quite large. However, since a given loop integral 
generically produces terms $a=\lambda/(16\pi^2)$, one must examine explicit
higher-loop results to judge whether this two-loop zero is a robust,
reliable prediction of perturbation theory or whether, on the contrary, it 
occurs at too large a value of $\lambda$ to be a reliable prediction.  We 
address this question here. 

% =====================================================================

\subsection{General Methods for Analysis of Zeros of $\beta_{a,n\ell}$ 
for $n \ge 3$} 

We proceed to calculate zeros of the $n$-loop beta function
$\beta_{a,n\ell}$ for $n \ge 3$.  In general, the condition that the
$n$-loop beta function $\beta_{a,n\ell}$ has a zero away from the origin
$a=0$ is the polynomial equation of degree $n-1$ in $a$:
\beq
\sum_{\ell=1}^n b_n a^{\ell-1} = 0 \ . 
\label{beta_nloop_reduced_zero}
\eeq
Although only one of the roots of Eq. (\ref{beta_nloop_reduced_zero}) will be
relevant for our analysis, it will be useful to characterize the full set of
roots, as in \cite{bc}. To do this, one may make use of information from the 
discriminant of Eq. (\ref{beta_nloop_reduced_zero}), denoted 
$\Delta_n(b_1,b_2,...,b_{n+1})$. Some relevant formulas on
discriminants are given in Appendix \ref{discriminants}. 

% ========================================================================

\subsection{ $\beta_{a,3\ell}$}

The condition that the three-loop beta function, $\beta_{a,3\ell}$, vanishes at
a nonzero value of $a$, in addition to the IR zero at $a=0$, is the special
case of Eq. (\ref{beta_nloop_reduced_zero}) with $n=3$, viz., the quadratic
equation $b_1+b_2a +b_3a^2=0$.  We find that for $b_3$ calculated in the
$\overline{\rm MS}$ scheme, this equation has no physical solutions.  Formally,
the solutions of this quadratic equation are
\beq
a = \frac{1}{2b_3}\Big (-b_2 \pm  \sqrt{\Delta_2(b_1,b_2,b_3)} \ \Big ) \ . 
\label{asol_3loop}
\eeq
However, these expressions are complex, as is evident from the fact that the
discriminant is negative (for all physical values of $N$):
\begin{widetext}
\beqs
\Delta_2(b_1,b_2,b_3) 
& = & -\frac{1}{81}\Big ( \frac{33}{2}N^3 + 512N^2+4412N+10076 \Big ) 
- \frac{16}{27}\Big ( 5N^2 + 62N + 176 \Big ) \zeta(3) \cr\cr
& = & - ( 0.2037N^3 + 9.8826N^2 + 98.6336N +249.7651)  \ . 
\label{delta2}
\eeqs
\end{widetext}
Thus, with $b_3$ calculated in the $\overline{MS}$ scheme, 
$\beta_{a,3\ell}$ does not have any physical UV zero. 

% =====================================================================

\subsection{ $\beta_{a,4\ell}$}

The condition that $\beta_{a,4\ell}=0$ for $a \ne 0$ is the $n=4$
special case of Eq. (\ref{beta_nloop_reduced_zero}), namely, the 
cubic equation
\beq
b_1+b_2a+b_3a^2+b_4a^3=0 \ .
\label{eqbeta_4loop_reduced}
\eeq
The nature of the roots of Eq. (\ref{eqbeta_4loop_reduced}) is determined by
the sign of the discriminant, 
\beqs
\Delta_3 \equiv \Delta_3(b_1,b_2,b_3,b_4) & = & b_2^2b_3^2-27b_1^2b_4^2 -
4(b_1b_3^3+b_4b_2^3) \cr\cr
& + & 18b_1b_2b_3b_4 \ .
\label{disc3}
\eeqs
The following properties of $\Delta_3$ will be useful here \cite{disc}: (i) if
$\Delta_3 > 0$, then all of the roots of Eq. (\ref{eqbeta_4loop_reduced}) are
real; (ii) if $\Delta_3 < 0$, then Eq.  (\ref{eqbeta_4loop_reduced}) has one
real root and a complex-conjugate pair of roots; (iii) if $\Delta_3=0$, then at
least two of the roots of Eq. (\ref{eqbeta_4loop_reduced}) coincide.  We find
that, with $b_3$ and $b_4$ calculated in the $\overline{\rm MS}$ scheme,
$\Delta_3(b_1,b_2,b_3,b_4)$ (which is a polynomial of degree 8 in $N$) is
negative for all physical $N$. Hence, the solutions to
Eq. (\ref{eqbeta_4loop_reduced}) consist of one real root and a
complex-conjugate pair of roots.  We display values of the real root 
$a_{_{UV,4\ell}}$ for
various values of $N$ in Table \ref{auv_nloop_values}.  We find that as $N$
increases from 1 to $N \simeq 770$, the real root decreases from 0.233 to
approximately 0.0714, but then increases again and diverges as $N \nearrow
N_{b4z} = 3218.755$, where $b_4$ vanishes.  For $N \ge N_{b4z}$,
$\beta_{a,4\ell}$ has no physical UV zero. (The positive real root that
diverged as $N \nearrow N_{b4z}$ now occurs at negative real values for this
range of $N$, and the other two roots of Eq. (\ref{eqbeta_4loop_reduced})
continue to be a complex-conjugate pair.)

% ========================================================================

\subsection{ $\beta_{a,5\ell}$}

The condition for a zero of $\beta_{a,5\ell}$ with $a \ne 0$ is the
special case of Eq. (\ref{beta_nloop_reduced_zero}) with $n=5$, namely, the
quartic equation
\beq
b_1+b_2a+b_3a^2+b_4a^3+b_5a^4=0 \ .
\label{beta_5loop_reduced_zero}
\eeq
The discriminant, $\Delta_4 \equiv \Delta_4(b_1,b_2,b_3,b_4,b_5)$, of this
equation is given by Eqs. (\ref{deltam4}) and (\ref{cbrel}) in Appendix
\ref{discriminants}.  We have calculated this and found that it is positive for
all physical $N$ except for the interval
\beq
493.096 < N < 504.740
\label{ndelta4negative}
\eeq
We denote the lower end of this interval as $N_{\Delta_4 z}=493.096$. For the
interval of $N$ from 1 to $N_{\Delta_4 z}$, Eq. (\ref{beta_5loop_reduced_zero})
has no physical solutions.  As $N$ (analytically continued from the positive
integers to the reals) increases through the value $N_{\Delta_4 z}$, a double
real root of Eq. (\ref{beta_5loop_reduced_zero}) appears at $a=0.1264$ and
then bifurcates into two real roots.  The smaller of these is the physical
$a_{_{UV,5\ell}}$ and decreases below 0.1264 as $N$ increases beyond
$N_{\Delta_4 z}$, while the larger root increases above $a=0.1264$ as $N$
increases above $N_{\Delta_4 z}$. As $N$ (again, considered as a real variable)
approaches the value $N=504.740$ from below, the larger real root diverges,
leaving only the lower one.  This continues to decrease as $N$ increases
further.  We list values of $a_{_{UV,5\ell}}$ for various $N$ in Table
\ref{auv_nloop_values}.

% =======================================================================

\subsection{Comparison of Calculations to Different Loop Orders}

A necessary condition for a perturbative calculation of the beta function 
$\beta_a$ to be reliable is that the fractional change
\beq
\bigg | \frac{\beta_{a,n+1}-\beta_{a,n}}{\beta_{a,n}} \bigg | 
\label{betachange}
\eeq
should generally decrease as the loop order $n$ increases, at least away from a
zero of $\beta_{a,n}$.  Another necessary condition for the reliability of a
result on a zero of the $n$-loop beta function, $\beta_{a,n}$, is that when one
calculates the beta function to the next higher-loop order, viz.,
$\beta_{a,n+1}$, the zero should still be present and its value should not
shift very much.  For the specific case at hand, where we are investigating a
possible UV zero of $\beta_a$, this condition is that the fractional shift
\beq
\frac{|a_{_{UV,n+1}}-a_{_{UV,n}}|}{a_{_{UV,n}}}
\label{achange}
\eeq
should be small.  Our calculations above show that neither of these two
necessary conditions is satisfied for this theory.  As was evident from our
plots of the $n$-loop beta functions and the ratios $R_n$ of the $n$-loop beta
function divided by the one-loop beta function given above, the fractional
change (\ref{betachange}) is not small for the values of $a$ that are relevant
for the analysis of a possible UV zero, even for the highest loop order $n$
that we have investigated. Recall from Eqs. (\ref{auv_2loop})-
(\ref{auv_2loop_Neqinfty}) that the values of $a_{_{UV,2\ell}}$ range from 9/17
to 1/3 as $N$ increases from 1 to $\infty$. Furthermore, although the two-loop
beta function $\beta_{a,2\ell}$ exhibits a UV zero, this is absent in the
three-loop beta function $\beta_{a,3\ell}$ and although the four-loop and five
loop beta functions have UV zeros for certain ranges of $N$, they 
occur at rather different values than for $\beta_{a,2\ell}$.  For example, for
$N=1$, $a_{_{UV,4\ell}}=0.233$, which is substantially smaller than
$a_{_{UV,2\ell}}=0.529$, and neither the three-loop nor five-loop beta function
has a UV zero.  A similar situation holds for $N=100$.  For $N=1000$,
$a_{_{UV,4\ell}}=0.0724$, which again is considerably smaller than
$a_{_{UV,2\ell}}=0.334$, and $\beta_{a,3\ell}$ has no UV zero, while
$a_{_{UV,5\ell}}=0.0228$, which is a substantially different value than both
the two-loop and four-loop UV zeros.  Similar comments apply for other values
of $N$.  Our higher-loop analysis therefore leads us to conclude that the 
(perturbatively calculated) beta function of this theory does not exhibit a 
robust, reliably calculable UV zero to the highest loop order, namely five
loops, to which it has been computed. 

% ====================================================================

\section{Large-$N$ Limit}

Further insight into the question of a UV zero of the beta function of this
theory can be obtained from an analysis of the limit
\beq
N \to \infty \ , \ \ {\rm with} \ x(\mu) \equiv Na(\mu) \ {\rm a \ finite \
  function \ of} \ \mu. 
\label{ln}
\eeq
We denote this as the LN limit and will use the symbol $\lim_{LN}$ to refer to
it. For the purpose of this analysis, we define a rescaled beta function that
is finite in the LN limit.  For large $N$, the two scheme-independent
coefficients have the asymptotic behavior $b_1 \sim N/3$ and $b_2 \sim -N$,
while $b_\ell \sim {\rm const.} \times N^{\ell-1}$ for $\ell \ge 3$ for the
higher-loop coefficients that have been calculated in the $\overline{\rm MS}$
scheme. From Eq. (\ref{b1}) one can write
\beq
b_1 = b_{1,1}N + b_{1,0} \quad {\rm where} \ b_{1,1}=\frac{1}{3}, \quad
b_{1,0} = \frac{8}{3} \ . 
\label{b1structure}
\eeq
We thus extract the leading-$N$ factors and define 
\beq
\check b_\ell =\lim_{LN} \frac{b_\ell}{N^{\ell-1}} \quad {\rm for} \ \ell 
\ge 2 \ . 
\label{bellhat}
\eeq
so that these $\check b_\ell$ are finite in the large-$N$ limit. 
The explicit values of the $\check b_\ell$ follow from the expressions given
above for the $b_\ell$; thus, $\check b_2 = -1$, $\check b_3 = 11/72$, etc.
Since the LN limit is defined so that $x(\mu)$ is a finite function of
$\mu$, the appropriate beta function that is finite in this limit is 
\beqs
\beta_x & = & \frac{dx}{dt} = \lim_{LN} N \beta_a \cr\cr
& = & x^2 \bigg [ b_{1,1} + \frac{1}{N} \sum_{\ell=2}^\infty \check
b_\ell \, x^{\ell-1} \bigg ] \ . 
\label{betax}
\eeqs

The $n$-loop beta function in the LN limit, denoted $\beta_{x,n\ell}$, is
defined via Eq. (\ref{betax}) with the upper limit on the sum being $\ell=n$
rather than $\ell=\infty$.  From Eq. (\ref{betax}), it is evident that in the
LN limit, for any given loop order $n$, $\beta_{x,n\ell}$ has no UV zero
$x_{_{UV,n\ell}}$, since
\beq
\lim_{LN} \frac{1}{N} \sum_{\ell=2}^n 
\Big ( \frac{\check b_\ell}{b_{1,1}} \Big ) \, x^{\ell-1} = 0 \ .
\label{lnsumzero}
\eeq
Hence, in the $N \to \infty$ limit, as $\mu$ increases, $x(\mu)$ increases,
eventually exceeding the range of values where the perturbative 
$n$-loop expansion of $\beta_{x,n\ell}$ is reliable.  This result in the 
LN limit agrees with our specific calculations for large finite values of $N$
as shown in Table \ref{auv_nloop_values}.

% =====================================================================

\section{Effect of Scheme Transformations}

In view of the fact that the $b_\ell$ with $\ell \ge 3$ are scheme-dependent,
one is motivated to study the effect of a scheme transformation on the beta
function of this $\lambda |\vec{\phi}|^4_4$ theory.  It should be recalled that
scheme dependence is also present, e.g., in higher-loop perturbative
calculations in quantum chromodynamics (QCD) and does not prevent one from
using such calculations successfully in comparisons with data
\cite{brodskyreview}.  If one were interested in a zero of the beta function at
zero coupling, as with the UV fixed point of an asymptotically free gauge
theory like QCD \cite{thooft77} or the IR fixed point of the $\lambda
|\vec{\phi}|^4_4$ theory, then one would expect that it should be possible to
transform away the terms in the beta function at loop order $\ell \ge 3$, and
an explicit construction that does this was presented in \cite{sch}.  However,
it was also pointed out in \cite{sch} that it is considerably more difficult to
construct an acceptable scheme transformation that removes some set of
coefficients at loop order 3 or higher at a zero of the beta function away from
the origin in coupling constant space than it is at the origin. Ref. \cite{sch}
gave a set of conditions that such a scheme transformation must satisfy to be
physically acceptable.  Here we recall the basic formalism; we refer the reader
to \cite{sch}-\cite{sch3} for further details (see also \cite{kataev_scheme}).

A scheme transformation can be expressed as a mapping between $\lambda$ and
$\lambda'$, or equivalently, $a$ and $a'$, which we write as $a = a' f(a')$. We
will refer to $f(a')$ as the scheme transformation function.  Since scheme
transformations cannot change the theory in the limit where the coupling goes
to zero, one requires that $f(0) = 1$.  We expand $f(a')$ as a power series of
the form
\beq
f(a') = 1 + \sum_{s=1}^{s_{max}} k_s \, (a')^s \ , 
\label{fxprime}
\eeq
where the $k_s$ are constants, and, {\it a priori}, $s_{max}$ may be finite or
infinite.  After the scheme transformation is applied, the beta function in the
new scheme has the form (\ref{betaa}) with $a$ replaced by $a'$ and $b_\ell$
replaced by $b'_\ell$.  Expressions for the $b_\ell'$ in terms of the $b_\ell$
and $k_s$ were given in \cite{sch}.  Ref. \cite{sch3} presented a generalized
one-parameter family of scheme transformation denoted $S_{R,m,k_1}$ with $m \ge
2$ and $s_{max}=m$ that can render $b_\ell'=0$ for $3 \le \ell \le m+1$
inclusive and can be applied at a zero of the beta function away from the
origin.

A natural approach is to investigate the effect of applying scheme
transformations in the one-parameter family $S_{R,2,k_1}$
to the beta function in the $\overline{\rm MS}$ scheme in order to 
render $b_3'=0$ in the transformed scheme.  This family of scheme 
transformations depends on a parameter $k_1$ and has 
\beq
k_2 = \frac{b_3}{b_1} + \frac{b_2}{b_1} \, k_1 + k_1^2  
\label{k2solk1}
\eeq
(with $k_s = 0$ for $s \ge 3$). If this scheme 
transformation were to be applicable, then in the transformed scheme the beta
function would have the form, for $n=3$, 
\beq
\beta_{a',3\ell} = (a')^2 \Big [ b_1 + b_2 \, a' \Big ] 
\label{betaprime_3loop}
\eeq
and, for $n \ge 4$, 
\beq
\beta_{a',n\ell} = (a')^2 \Big [ b_1 + b_2 \, a'  + \sum_{\ell=4}^n b_\ell'
  \, (a')^{\ell-1} \Big ] \quad {\rm for} \ n \ge 4 \ . 
\label{betaprime_nloop}
\eeq

One of the necessary conditions for the acceptability of
the scheme transformation is that $f(a')$ must be positive and not too
different from unity, since otherwise the transformation or its inverse would
map a reasonably small value of the coupling, for which perturbative methods
could be reliable, to an excessively large value, beyond the region where these
methods could be reliably applied.  In particular, this condition must be
satisfied at (and in the neighborhood of) the scheme-independent value
$a=a_{_{UV,2\ell}}=a'_{_{UV,2\ell}}$. We thus consider the 
evaluation of $f(a')$ at this point. We have
\beq
f(a'_{_{UV,2\ell}}) = 1 + \frac{b_1b_3}{b_2^2}
+ \frac{b_1^2}{b_2^2} k_1^2 \ ,
\label{fap_sr2k1_air2loop}
\eeq
where $b_3$ is calculated in the $\overline{\rm MS}$ scheme.  In addition to
the first term, both the second and third terms on the right-hand side of
Eq. (\ref{fap_sr2k1_air2loop}) are positive in the $\overline{\rm MS}$ scheme,
since $b_3$ is positive. Given that the coefficient of $k_1^2$ is positive, it
follows that $f(a'_{_{UV,2\ell}})$ is minimized by taking $k_1=0$.  We thus
choose $k_1=0$, so that $S_{R,2,k_1}$ reduces to the $S_{R,2}$ transformation
\cite{sch}-\cite{sch3}. Even with this choice, we find that as $N$ increases,
$f(a'_{_{UV,2\ell}})$ quickly becomes excessively large, preventing one from
using this scheme transformation over a very large interval of values of $N$.
A second condition is that the Jacobian $da/da'$ should not approach or equal
zero, since otherwise the scheme transformation is singular. As with $f(a')$,
we require this condition to be satisfied at (and in the neighborhood of)
$a=a_{_{UV,2\ell}}=a'_{_{UV,2\ell}}$. At this point,
\beq
J = 1 + \frac{3b_1b_3}{b_2^2} + \frac{b_1}{b_2}k_1
+ \frac{3b_1^2}{b_2^2}k_1^2 \ .
\label{j_sr2k1_air2loop}
\eeq
Since the term $3b_1b_3/b_2^2$ is positive, the choice that we have made, 
namely $k_1=0$, also guarantees that $J$ stays positive. 

By construction, after application of the $S_{R,2,k_1}$ scheme transformation,
the three-loop beta function $\beta_{a',3\ell}$, has a UV zero, in contrast to
the situation in the original $\overline{\rm MS}$ scheme, where it does not.
This is evident from the $n=3$ special case of Eq. (\ref{betaprime_nloop}).
Since $b_3'=0$, this UV zero occurs at the same value of $a'$ as the two-loop
value, i.e.,
\beq
a'_{_{UV,3\ell}} = a'_{_{UV,2\ell}} = a_{_{UV,2\ell}} = -\frac{b_1}{b_2} \ . 
\label{auv_3loop}
\eeq
However, in order for one to take this as a significant indication that there
is, in fact, a UV zero in the beta function, it must continue to be present
when calculated to higher order (at least for the range of $N$ where this
$S_{R,2}$ scheme transformation can be applied) and at higher orders, the
fractional change in the value should decrease to be reasonably small values.

To investigate whether these conditions are satisfied, we have calculated the
four-loop and five-loop beta functions in the $S_{R,2}$-transformed scheme,
$\beta_{a',4\ell}$ and $\beta_{a',5\ell}$ and have investigated their zeros.
For this calculation we use the expressions for $b_4'$ and $b_5'$ from
\cite{sch}-\cite{sch3}.  A UV zero of the four-loop beta function in the
transformed scheme, $\beta_{a',4\ell}$, is given by the relevant root of the
cubic equation $b_1+b_2 \, a'+b'_4 \, (a')^3=0$, namely the positive root
nearest to the origin.  Similarly, a UV zero of $\beta_{a',4\ell}$, would be
given by the relevant root of the quartic equation $b_1+b_2 \, a'+b'_4 \,
(a')^3 + b'_5 \, (a')^4 =0$, namely a positive root nearest to the origin.  We
show the results in Table \ref{sr2results} for the range of $N$ where
$f(a'_{_{UV,2\ell}})$ is not excessively large, namely $1 \le N \lsim 10$.  For
illustration, we also show calculations for the value $N=100$, although $f(a')$
is arguably too large for the scheme transformation to be applicable at this
value of $N$.  As is evident from Table \ref{sr2results}, although
$\beta_{a',4\ell}$ has a UV zero, it occurs at a considerably smaller value of
$a'$ than the two-loop value, $a'_{_{UV,2\ell}}$.  Furthermore, we find that at
five-loop order the beta function in the transformed scheme,
$\beta_{a',5\ell}$, does not have a (physical) UV zero.  (It has two pairs of
complex-conjugate roots). Therefore, over the interval of $N$ where the
$S_{R,2}$ scheme transformation can be applied without excessively large
$f(a')$, the necessary conditions stated above for these results to be
consistent with a robust UV zero of the beta function are not satisfied; i.e.,
although $\beta_{a',4\ell}$ has a UV zero, it occurs at a considerably smaller
value of $a'$ than $a'_{_{UV,2\ell}}$ and, furthermore, $\beta_{a',5\ell}$ does
not have a UV zero.  By continuity, our results also apply for an (infinite)
set of other $S_{R,2,k_1}$ scheme transformations whose functions $f(a')$ are
close to the function $f(a')$ for $S_{R,2}$, namely the set with small $|k_1|$.

Our results from the analysis of the beta function up to five-loop order after
application of the $S_{R,2,k_1}$ scheme transformations thus agree with our
results from the analysis in the original $\overline{\rm MS}$ scheme; in both
studies, we do not find evidence that the beta function of this theory has a UV
zero, at least insofar as we can use perturbative methods reliably to
investigate it.  Indeed, this was already clear from
Figs. \ref{lam_beta_Neq1}-\ref{lam_betareduced_Neq100}. These showed that
although the two-loop beta function has a UV zero, this occurs at a value
$a_{_{UV,2\ell}}$ (dependent on $N$) that is well beyond the range where the
perturbative calculation is reliable, since the respective higher-loop beta
functions $\beta_{a,n\ell}$ with $n=3, \ 4, \ 5$ loops differ considerably from
$\beta_{a,2\ell}$ for $a \simeq a_{_{UV,2\ell}}$.  The absence of a reliably
calculable UV zero of $\beta_{a,n\ell}$ means that, to the highest loop order,
namely $n=5$ loops, to which the beta function has been calculated,
$\lambda(\mu)$ increases with increasing $\mu$, eventually exceeding the range
where perturbative methods of analysis can be used.  Some recent discussions of
possibilities for the nonperturbative behavior of $\lambda |\vec{\phi}|^4_4$
theory include \cite{zinnjustinbook,recentnonpert}.

A scheme transformation constitutes one type of resummation of a perturbation
series.  A different method of analysis is provided by Pad\'e approximants
\cite{pade1}.  We calculate and analyze these next.

% ======================================================================== 

\section{Analysis With Pad\'e Approximants}

Given a series for an abstract function $f(z) = \sum_{s=0}^m c_s z^s$,
the $[p,q]$ Pad\'e approximant is the rational function 
\beq
[p,q] = \frac{\sum_{j=0}^p N_j z^j}{\sum_{k=0}^q D_k z^k}
\label{pqpade}
\eeq
with polynomials in the numerator and denominator of degree $p$ and $q$,
respectively, where $p+q=m$ \cite{pade1}.  Without loss of generality,
one may take $D_0=1$, so that $N_0=c_0$.  The coefficients $N_j$ with
$j=1,...,p$ and $D_k$ with $k=1,...,q$ are determined by the $m$ coefficients
$c_1,...,c_m$.  Thus, a $[p,q]$ Pad\'e approximant to $f(z)$ is a closed-form
rational approximation whose Taylor series expansion in $z$ matches the power
series for $f(z)$ to the highest order to which it is calculated. 

It is natural to inquire how the zeros of various Pad\'e approximants to
$\beta_{a,n\ell}$ compare with those of $\beta_{a,n\ell}$, as we did, e.g., in
\cite{bvh} for the IR zero of the beta function of a non-Abelian gauge theory.
For this purpose, it is convenient to extract the overall factor of $a^2$ in
$\beta_{a,n\ell}$ and thus compute the Pad\'e approximants to the function
$\beta_{a,n\ell}/a^2 = b_1 + \sum_{\ell=2}^n b_\ell a^{\ell-1}$.  At a given
loop order $n$, we can calculate the $[p,q]$ Pad\'e approximants with
$p+q=n-1$.  The $[n-1,0]$ Pad\'e approximant to $\beta_{a,n\ell}/a^2$ is the
function itself, and the $[0,n-1]$ approximant has no zeros so we do not
consider these.  The use of Pad\'e approximants to scattering amplitudes has a
long history in particle physics and, as usual, the fact that the Pad\'e
approximant has a Taylor series expansion with a finite radius of convergence
is not to be taken as implying that the actual function being approximated (in
this case, $\beta_{n\ell}/a^2$) has such a Taylor series expansion in powers of
the coupling; indeed, it is known that the series expansion (\ref{betaa}) is
only an asymptotic expansion \cite{asymptotic,sm}. Clearly, there are several
necessary conditions for a zero of a $[p,q]$ Pad\'e approximant to
$\beta_{n\ell}/a^2$ to be taken to be physically meaningful. Two of these
conditions are (i) that this zero must
occur on the positive real axis in the complex $a$ plane at a value that is not
too different from $a_{_{UV,2\ell}}$; (ii) the location of the zero must be
closer to the origin $a=0$ than any of the $q$ poles. Moreover, no implication
is made that these approximants accurately describe the large-$a$ behavior of 
the beta function (which would be $\beta_{n,\ell} \propto a^{2+p-q}$). 

We have carried out this analysis with Pad\'e approximants and have found that
it confirms the conclusions that we reached from our study of the $n$-loop beta
function $\beta_{a,n\ell}$ itself.  For example, let us consider the case
$N=1$.  As with other values of $N$, the three-loop beta function,
$\beta_{a,3\ell}$, has no UV zero. (Aside from the double zero at $a=0$, it has
a complex-conjugate pair of zeros at $a=0.08705 \pm 0.29084i$.) Similarly, the
relevant Pad\'e approximant to $\beta_{a,3\ell}/a^2$, namely [1,1], has only an
unphysical zero at $a=-0.2594$ (as well as a pole at an unphysical point even
closer to the origin, namely $a=-0.19745$).  At the four-loop level,
$\beta_{a,4\ell}$ has a UV zero at $a=0.2333$, but this is not reproduced by
either of the relevant Pad\'e approximants to $\beta_{a,4\ell}/a^2$; the [1,2]
approximant has an unphysical zero at $a=-0.1294$ (and unphysical poles at
$a=-0.1138$ and $a=-1.2005$), while the [2,1] approximant has an unphysical
zero at $a=-0.1400$ and a UV zero at $a=1.4543$, much larger than the UV zero
of $\beta_{a,4\ell}$ (as well as a pole nearer to the origin, at the unphysical
value $a=-0.1198$).

 At the five-loop level, $\beta_{a,5\ell}$ has no physical UV zero. (In
addition to the double zero at $a=0$, it has zeros at the two complex-conjugate
pairs $a=-0.09440 \pm 0.14585i$ and $a=0.1421 \pm 0.1213i$).  Of the three
relevant Pad\'e approximants to $\beta_{a,5\ell}/a^2$, the [1,3] approximant
has a zero at $a=-0.0949$ (and poles at $a=-0.0899$, $a=-0.4644$, and
$a=1.1714$); [2,2] has zeros at $a=-0.0874$ and $a=-0.5298$ (and poles at
$a=-0.0840$ and $a=-0.3013$), and [3,1] has zeros at $a=-0.10245$ and
$a=0.2439 \pm 0.6002i$ (and a pole at $a=-0.09535$).  Evidently, all of the
zeros of these Pade\'e approximants are unphysical.

We find similar results for other values of $N$. Thus, from our calculation and
analysis of Pad\'e approximants to the $n$-loop beta function, we add to our
evidence against a stable, reliably calculable UV zero in the $\lambda
|\vec{\phi}|^4_4$ theory.  Other resummation methods such as a Borel transform
could also be applied, but the findings from these two methods, namely scheme
transformations and Pad\'e approximants already provide strong evidence against
a robust UV zero in the beta function for this theory up to the five-loop
order.

% =======================================================================

\section{Discussion}

In this section we compare our present results with those on zeros of the beta
function for some other theories.  We begin with the nonlinear $\sigma$ model
in $d=2+\epsilon$ dimensions \cite{nlsm}.  Both the $\lambda |\vec{\phi}|^4_4$
theory and the nonlinear $\sigma$ model in $d=2+\epsilon$ are IR-free, but in
the NL$\sigma$M one can choose a parameter, namely $\epsilon$, to approach zero
so as to make the UV fixed point occur at an arbitrarily weak coupling. In this
NL$\sigma$M let us define appropriately rescaled quantities $\tilde \lambda
\equiv \lambda N$ and $\beta_{\tilde \lambda} = d{\tilde\lambda}/dt$, so that
Eq. (\ref{beta_nlsm}) with Eq.  (\ref{lambda_c_nlsm}) reads
\beq
\beta_{\tilde \lambda} = \epsilon \tilde \lambda
\Big ( 1 - \frac{\tilde \lambda}{\tilde \lambda_c} \Big ) \ , 
\label{beta_nlsm_rescaled}
\eeq
where $\tilde \lambda_c = 2\pi \epsilon$.  As noted before, this result was
obtained by summing an infinite set of Feynman diagrams that dominate in the
large-$N$ limit.  By letting $\epsilon \to 0^+$, one can
make the UVFP occur at an arbitrarily small value of
$\tilde \lambda_c$. 

A different but related comparison can be made with the calculation of the IR
zero of the beta function for the gauge coupling, $\beta_\alpha = d\alpha/dt$
(where $\alpha = g^2/(4\pi)$), that is present in an asymptotically free
vectorial non-Abelian gauge theory in $d=4$ dimensions with gauge group $G$ and
an appropriately large fermion content.  Thus, consider such a theory with
$N_f$ copies of massless fermions transforming according to a representation
$R$ of $G$ and denote $N_{f,b1z}$ as the upper bound on $N_f$ for asymptotic
freedom and $N_{f,b2z}$ as the value of $N_f$ such that for $N_f > N_{f,b2z}$,
the one-loop and two-loop terms in $\beta_\alpha$ have opposite sign.  Since
$N_{f,b2z} < N_{f,b1z}$, it follows that there is an interval in $N_f$, namely
$N_{f,b2z} < N_f < N_{f,b1z}$ for which the two-loop $\beta_\alpha$ function
has an IR zero.  Importantly, all of the $n$-loop beta functions for $n=2, \ 3,
\ 4$ loop orders consistently exhibit an IR zero, and the fractional shift in
this IR zero is reduced when one compares the beta functions at $n=3$ and $n=4$
loop order, as contrasted with the shift between the beta functions at $n=2$
and $n=3$ loop order.  This is evident from Figs. 1 and 2 of Ref. \cite{bc}.
Furthermore, with appropriate choices of $G$, $R$, and $N_f$, one can make this
IR zero occur at a small value of $\alpha$ \cite{bz}.  In particular, consider
an asymptotically free vectorial gauge theory with $G={\rm SU}(N_c)$, $R$ equal
to the fundamental representation, $N_c \to \infty$, $N_f \to \infty$, with
$r=N_f/N_c$ fixed, and $\xi(\mu) \equiv \alpha(\mu)^2 N_c$ a finite function
of $\mu$. Then for $r$ in the interval $34/13 < r < 11/2$, the rescaled beta
function $\beta_\xi = d\xi/dt$ has an IR zero at the two-loop level at
\beq
\xi_{UV,2\ell} = \frac{4\pi(11-2r)}{13r-34} \ . 
\label{xiuv_2loop_lnn}
\eeq
The value of $\xi_{UV,2\ell}$ can be made arbitrarily small by letting $r$
approach 11/2 from below.  The situation in the $\lambda |\vec{\phi}|^4_4$
theory is fundamentally different from that in this non-Abelian gauge theory
because there is no parameter that can be tuned to make the two-loop value of
$a_{_{UV,2\ell}} \ll 1$ or, in the large-$N$ limit, the value of
$x_{_{UV,2\ell}} \ll 1$.  This is similar to the situation in the U(1) gauge
theory in $d=4$ with $N_f$ copies of a charged fermion, where we found evidence
against a UVFP in \cite{lnf} (see also \cite{holdom}).

In the present work we have focused on a simple O($N$) scalar field theory
involving a single coupling, using calculations to the rather high order of
five loops.  This is complementary to studies of more complicated theories with
more than one coupling, involving RG calculations to lower than five-loop
order.  In passing, we add a few comments on these theories. The RG behaviors
of theories with scalar and fermion fields have been studied for many years,
both perturbatively and nonperturbatively (for references to the literature,
see, e.g., \cite{srgt}).  For small values of the quartic scalar coupling
$\lambda$ and the Yukawa coupling $y$, both $\beta_\lambda = d\lambda/dt$ and
$\beta_y = dy/dt$ are positive, so that near the origin $(\lambda,y) = (0,0)$,
as $\mu$ decreases, the RG flow is toward to origin, i.e. a free theory.  One
may investigate such theories for possible fixed points of the renormalization
group away from the origin. For sufficiently large couplings, one must use
nonperturbative methods, such as lattice simulations \cite{yuk}. Studies have
also been performed of models with gauge, fermion, and scalar fields, and hence
three or more couplings \cite{cel,cp3}.  In some cases, UV-stable fixed points
have been reported (e.g., \cite{cp3} and references therein), motivating
continued interest in the phenomenon of asymptotic safety.  For the actual
Standard Model, an intriguing feature is that because of the the large
top-quark Yukawa coupling, instead of increasing for large $\mu$ in the
ultraviolet, the quartic Higgs coupling actually decreases and eventually
vanishes \cite{strumia}, although this is sensitive to ultraviolet completions
of the Standard Model.

% ========================================================================

\section{Conclusions} 

In this paper we have investigated whether there is a reliably calculable
ultraviolet zero of the beta function in the O($N$) $\lambda |\vec{\phi}|^4_4$
field theory. This question is of interest since the two-loop beta function
does have a UV zero. We have examined whether two necessary conditions are
met, namely that the existence of this UV zero persists to higher-loop order
and that higher-order $n$-loop calculations yield reasonably stable values of
$a_{_{UV,n\ell}}=\lambda_{_{UV,n\ell}}/(16\pi^2)$.  We have carried out this
study using calculations of the $n$-loop beta functions $\beta_{a,n\ell}$ for
three-, four-, and five-loop order in the $\overline{\rm MS}$ scheme.  We have
shown that in this scheme, (i) $\beta_{\lambda,3\ell}$ has no UV zero; (ii)
$\beta_{\lambda,4\ell}$ has a UV zero only for a limited range of $N$, and has
no UV zero for sufficiently large $N$; (iii) $\beta_{\lambda,5\ell}$ has no UV
zero for $N$ from 1 to almost 500, and although it does have a UV zero for
larger values of $N$, this zero occurs at quite a different value of $a$ than
two-loop UV zero.  Thus, we find that neither of the two necessary conditions
for a robust, reliably calculable UV zero of the beta function is satisfied
for any $N$.  This inference is confirmed by our study of the effect of
applying scheme transformations to the beta function and from calculations of
Pad\'e approximants. Evidently, the zero in $\beta_{a,2\ell}$ occurs at too
large a value for the two-loop perturbative calculation to be accurate.  We
thus conclude that in the range of quartic coupling $\lambda$, or equivalently,
$a$, where the perturbative calculation of $\beta_{\lambda,n\ell}$ is reliable,
the O($N$) $\lambda |\vec{\phi}|^4_4$ theory does not exhibit robust evidence
of a UV zero up to the level of $n=5$ loops for any $N$. This conclusion is in
accord with the current view of the O($N$) $\lambda |\vec{\phi}|^4$ theory in
$d=4$ dimensions as an effective field theory, to be applied only over a
restricted range of momentum scales $\mu$.  

% =======================================================================

\begin{acknowledgments}

This research was partly supported by NSF grant PHY-13-16617. 

\end{acknowledgments}

% ================================================================

\begin{appendix}

\section{Discriminants} 
\label{discriminants}

Our study of the zeros of the $n$-loop beta function of the $\lambda
|\vec{\phi}|^4_4$ theory requires an analysis of the zeros of the polynomial
equation (\ref{beta_nloop_reduced_zero}), of degree $n-1$ in the variable $a$
given by Eq. (\ref{a}). At the loop order $\ell \ge 3$, this analysis is
considerably expedited by the use of the corresponding discriminant.

Consider the polynomial of degree $m$ in an abstract variable $z$,
\beq
P_m(z) = \sum_{s=0}^m c_s z^s 
\label{pmz}
\eeq
and denote the set of $m$ roots of the equation $P_m(z)=0$ as $\{z_1,...,z_m 
\}$.  The discriminant of this equation is defined as \cite{disc}
\beq
\Delta_m \equiv \Big [ c_m^{m-1} \prod_{i < j} (z_i-z_j) \Big ]^2 \ .
\label{deltam}
\eeq
Since $\Delta_m$ is a symmetric polynomial in the roots of the equation
$P_m(z)=0$ (being proportional to the square of the Vandermonde polynomial of
these roots), the symmetric function theorem implies that it can be expressed
as a polynomial in the coefficients of $P_m(z)$ \cite{symfun}. We will
sometimes indicate this dependence explicitly, writing
$\Delta_m(c_0,...,c_m)$. The discriminant $\Delta_m$ is a homogeneous
polynomial of degree $m(m-1)$ in the roots $\{ z_i \}$.  For the application to
the analysis of the roots of Eq. (\ref{beta_nloop_reduced_zero}), the
correspondence is $z=a$ and
\beq
c_s = b_{s+1} \ , \quad s=0,...,m \ . 
\label{cbrel}
\eeq
To analyze the zeros of $\beta_{a, n\ell}$ away from
the origin, given by the roots of Eq. (\ref{beta_nloop_reduced_zero}), of
degree $m=n-1$, we will thus use the discriminant
$\Delta_{n-1}(b_1,b_2,...,b_n)$. Because of the homogeneity properties of this
discriminant,
\beq
\Delta_{n-1}(\bar b_1,\bar b_2,...,\bar b_n) =
(4\pi)^{-(n+1)(n-2)}\Delta_{n-1}(b_1,b_2,...,b_n) \ .
\label{deltambbar}
\eeq

The discriminant $\Delta_m$ is most
conveniently calculated in terms of the Sylvester matrix of $P(z)$ and
$dP(z)/dz$, equivalent to the resultant matrix, denoted $S_{P,P'}$, of
dimension $(2m-1) \times (2m-1)$ \cite{disc}: 
\beq
\Delta_m = (-1)^{m(m-1)/2}c_m^{-1}{\rm det}(S_{P,P'}) \ .
\label{Deltammatrix}
\eeq
For our analysis, we use $\Delta_m$ for $m=2, \ , 3, \ 4$. 
The $m=2$ discriminant, $\Delta_2$, is elementary: 
$\Delta_2=c_1^2-4c_0c_2$. For $m=3$, 
\beq
S_{P_3,P_3'} = \left( \begin{array}{ccccc}
c_3  & c_2  & c_1  & c_0  & 0   \\
 0   & c_3  & c_2  & c_1  & c_0 \\
3_3  & 2c_2 & c_1  &  0   &  0  \\
  0  & 3c_3 & 2c_2 & c_1  &  0  \\
  0  &   0  & 3c_2 & 2c_2 & c_1 \end{array} \right )
\label{sp3pp3}
\eeq
so that 
\beqs
\Delta_3(c_0,c_1,c_2,c_3) & = & c_1^2 c_2^2 - 27c_0^2 c_3^2
- 4(c_0 c_2^3 + c_3 c_1^3) \cr\cr
& + & 18 c_0 c_1 c_2 c_3 \ .
\label{deltam3}
\eeqs

For $m=4$, 
\beq
S_{P_4,P_4'} = \left( \begin{array}{ccccccc}
c_4  & c_3  & c_2  & c_1  & c_0  &  0   &  0  \\
 0   & c_4  & c_3  & c_2  & c_1  & c_0  &  0  \\
 0   &  0   & c_4  & c_3  & c_2  & c_1  & c_0 \\
4c_4 & 3c_3 & 2c_2 & c_1  &  0   &  0   &  0  \\
 0   & 4c_4 & 3c_3 & 2c_2 & c_1  &  0   &  0  \\
 0   &  0   & 4c_4 & 3c_3 & 2c_2 & c_1  &  0  \\
 0   &  0   &  0   & 4c_4 & 3c_3 & 2c_2 & c_1 \end{array} \right )
\label{sp4pp4}
\eeq
so that
\begin{widetext} 
\beqs
\Delta_4(c_0,c_1,c_2,c_3,c_4) & = & 
c_1^2c_2^2c_3^2 - 128c_0^2c_2^2c_4^2 - 4c_1^3c_3^3 + 256c_0^3c_4^3 
- 27(c_0^2c_3^4 + c_1^4c_4^2) - 4(c_0c_2^3c_3^2 + c_1^2c_2^3c_4) \cr\cr
& + & 18(c_0c_1c_2c_3^3 + c_1^3c_2c_3c_4) - 6c_0c_1^2c_3^2c_4
+ 144(c_0c_1^2c_2c_4^2+c_0^2c_2c_3^2c_4) + 16c_0c_2^4c_4 \cr\cr
& - & 192c_0^2c_1c_3c_4^2 - 80c_0c_1c_2^2c_3c_4  \ . 
\label{deltam4}
\eeqs
\end{widetext}

\end{appendix}

\bigskip
\bigskip
\bigskip

\newpage

% table I
\begin{table}
\caption{\footnotesize{Values of the $\bar b_n=b_n/(16\pi^2)^n$, where $b_n$
are the $n$-loop beta function coefficients in Eqs. (\ref{beta}) and
(\ref{betaa}), for $1 \le n \le 5$, as functions of $N$ for $1 \le N \le 20$
and selected larger values of $N$. The notation $r$e$n$ means $r \times 
10^n$.}}
\begin{center}
\begin{tabular}{|c|c|c|c|c|c|} \hline\hline
$N$ & $\bar b_1$ & $\bar b_2$ & $\bar b_3$ & $\bar b_4$ & $\bar b_5$ 
\\ \hline
1     & 0.2387   & $-0.03588$   & 0.01640   & $-0.01089$  & 0.09090   \\
2     & 0.2653   & $-0.04222$   & 0.02013   & $-0.01406$  & 0.01227   \\
3     & 0.2918   & $-0.04855$   & 0.02401   & $-0.01755$  & 0.01595   \\
4     & 0.3183   & $-0.05488$   & 0.02805   & $-0.02137$  & 0.02016   \\
5     & 0.3448   & $-0.06121$   & 0.03224   & $-0.02553$  & 0.02492   \\
6     & 0.3714   & $-0.06755$   & 0.03658   & $-0.03001$  & 0.03024   \\
7     & 0.3979   & $-0.07388$   & 0.04108   & $-0.03482$  & 0.03616   \\
8     & 0.4244   & $-0.08021$   & 0.04573   & $-0.03996$  & 0.04269   \\
9     & 0.4509   & $-0.08655$   & 0.05054   & $-0.04542$  & 0.04984   \\
10    & 0.4775   & $-0.09288$   & 0.05550   & $-0.05121$  & 0.05765   \\
100   & 2.8648   & $-0.6628$    &  1.1324   & $-1.87505$  & 5.4152    \\
200   & 5.5174   & $-1.2961$    &  3.7918   & $-6.7359$   & 26.0096   \\
300   &  8.1700  & $-1.9293$    &  7.9910   & $-14.2812$  & 54.2973   \\
400   & 10.8225  & $-2.5626$    & 13.7300   & $-24.2014$  & 63.0752   \\
500   & 13.4751  & $-3.1958$    & 21.0087   & $-36.1873$  &  5.4300   \\
600   & 16.1277  & $-3.8291$    & 29.8273   & $-49.9293$  & $-1.85262$e2 \\
700   & 18.7803  & $-4.4624$    & 40.1856   & $-65.1180$  & $-5.95335$e2  \\
800   & 21.4329  & $-5.0956$    & 52.0837   & $-81.4440$  & $-1.33083$e3 \\
1.0e3 & 26.7380  & $-6.3621$    & 80.4992   & $-1.16270$e2  & $-4.30084$e3 \\
2.0e3 & 53.2639  & $-12.6947$   & 3.149645e2& $-2.53435$e2 & $-1.01045$e5 \\
3.0e3 & 79.7897  & $-19.0273$   & 7.03408e2 & $-1.02078$e2 & $-5.63816$e5 \\
4.0e3 &1.063155e2 & $-25.3598$   & 1.24583e3 &   6.472275e2 & $-1.86330$e6 \\
\hline\hline
\end{tabular}
\end{center}
\label{btilde_nloop_values}
\end{table}
%

% table II
\begin{table}
\caption{\footnotesize{Values of the UV zero $a_{_{UV,n\ell}}$ of the $n$-loop
beta function, $\beta_{a,n\ell}$, for $n=2,...,5$, as a function of $N$,
with $b_n$, $n=3, \ 4, \ 5$ calculated in the $\overline{\rm MS}$
scheme. The dash notation $-$ means that $\beta_{a',n\ell}$ has no physical UV
zero. The special values of $N$ are 
$N_{\Delta_4 z}=493.096$ from Eq. (\ref{ndelta4negative}), 
$N_{b5z}=504.740$ from Eq. (\ref{Nb5z}), and 
$N_{b4z}=3218.755$ from Eq. (\ref{Nb4z}).}}
\begin{center}
\begin{tabular}{|c|c|c|c|c|c|} \hline\hline
$N$ & $a_{_{UV,2\ell}}$ & $a_{_{UV,3\ell}}$ & $a_{_{UV,4\ell}}$ & 
$a_{_{UV,5\ell}}$ 
\\ \hline
1        &  0.5294   & $-$    &  0.2333    & $-$      \\
2        &  0.5000   & $-$    &  0.2217    & $-$      \\
3        &  0.4783   & $-$    &  0.2123    & $-$      \\
4        &  0.4615   & $-$    &  0.2044    & $-$      \\
5        &  0.4483   & $-$    &  0.1978    & $-$      \\
6        &  0.4375   & $-$    &  0.1920    & $-$      \\
7        &  0.4286   & $-$    &  0.1869    & $-$      \\
8        &  0.42105  & $-$    &  0.1823    & $-$      \\
9        &  0.4146   & $-$    &  0.1783    & $-$      \\
10       &  0.4091   & $-$    &  0.1746    & $-$      \\
100      &  0.3439   & $-$    &  0.1012    & $-$      \\
$N_{\Delta_4 z}$&0.3356&$-$   &  0.07353   & 0.12636  \\ 
500      &  0.3355   &  $-$   &  0.07341   & 0.08045  \\
$N_{b5z}$&  0.3355   &  $-$   &  0.073325  & 0.073325 \\ 
1000     &  0.3344   &  $-$   &  0.07241   & 0.02276  \\
2000     &  0.3339   &  $-$   &  0.1054    & 0.01231  \\
3000     &  0.3337   &  $-$   &  0.5475    & 0.008850 \\
$N_{b4z}$&  0.3337   &  $-$   &  $-$       & 0.008366 \\
4000     &  0.3336   &  $-$   &  $-$       & 0.007042 \\
$10^4$   &  0.3334   &  $-$   &  $-$       & 0.003460 \\
\hline\hline
\end{tabular}
\end{center}
\label{auv_nloop_values}
\end{table}

% table III
\begin{table}
\caption{\footnotesize{Values of the transformation function
$f(a'_{_{UV,2\ell}})$ and resultant UV zeros $a'_{_{UV,n\ell}}$ of the $n$-loop
beta function, $\beta_{a',n\ell}$, for $n=2,...,5$, as a function of $N$,
resulting from the application of the $S_{R,2}$ scheme transformation to the
$\overline{\rm MS}$ scheme. The dash notation $-$ means that $\beta_{a,n\ell}$ 
has no physical UV zero.}}
\begin{center}
\begin{tabular}{|c|c|c|c|c|} \hline\hline
$N$ & $f(a'_{_{UV,2\ell}})$ & $a'_{_{UV,2\ell}}$, \ $n=2,3$ 
& $a'_{_{UV,4\ell}}$ & $a'_{_{UV,5\ell}}$ 
\\ \hline
1      &   4.0410   &  0.5294   &  0.1917   &  $-$   \\
2      &   3.9961   &  0.5000   &  0.1822   &  $-$   \\
3      &   3.9726   &  0.4783   &  0.1746   &  $-$   \\
4      &   3.9642   &  0.4615   &  0.1683   &  $-$   \\
5      &   3.9668   &  0.4483   &  0.1630   &  $-$   \\
6      &   3.9776   &  0.4375   &  0.1584   &  $-$   \\
7      &   3.9947   &  0.4286   &  0.1544   &  $-$   \\
8      &   4.0167   &  0.42105  &  0.1509   &  $-$   \\
9      &   4.0427   &  0.4146   &  0.1477   &  $-$   \\
10     &   4.0719   &  0.4091   &  0.1448   &  $-$   \\
100    &   8.3846   &  0.3439   &  0.08354  &  $-$   \\
\hline\hline
\end{tabular}
\end{center}
\label{sr2results}
\end{table}


\begin{thebibliography}{99}
\bibitem{alt}
%
Another common normalization uses ${\cal L}_{int} =
-(\lambda_{alt}/4)(\vec{\phi}^{\ 2})^2$. These are equivalent, with
$\lambda = 6\lambda_{alt.}$.
\bibitem{nreal}
%
Here and below, when expressions are given for $N$ that evaluate
to non-integral real values, it is understood that they are formal and are
interpreted via an analytic continuation of $N$ from physical nonnegative
integer values to real numbers.
\end{thebibliography}
\end{document}